\documentclass[conference]{IEEEtran}

\usepackage{graphicx}
\usepackage{amsmath}
\usepackage{cite}

\begin{document}
		

\graphicspath{{images/}}

\title{Simultaneous Acoustic Trapping and Imaging of Microbubbles at Clinically Relevant Flow Rates}

\author{\IEEEauthorblockN{Sevan~Harput, Luzhen~Nie, David~M.~J.~Cowell, Thomas Carpenter, Ben Raiton, \\ James~McLaughlan, and Steven~Freear} 

	 \IEEEauthorblockA{ \small Ultrasonics and Instrumentation Group, School of Electronic and Electrical Engineering, University of Leeds, Leeds, LS2 9JT, UK \\
	  	E-mail: S.Harput@leeds.ac.uk, S.Freear@leeds.ac.uk \\ 
	  	{ } }
}

\maketitle

\begin{abstract}

Mechanisms for non-invasive target drug delivery using microbubbles and ultrasound have attracted growing interest. Microbubbles can be loaded with a therapeutic payload and tracked via ultrasound imaging to selectively release their payload at ultrasound-targeted locations. In this study, an ultrasonic trapping method is proposed for simultaneously imaging and controlling the location of microbubbles in flow by using acoustic radiation force. Targeted drug delivery methods are expected to benefit from the use of the ultrasonic trap, since trapping will increase the MB concentration at a desired location in human body.

The ultrasonic trap was generated by using an ultrasound research system UARP II and a linear array transducer. The trap was designed asymmetrically to produces a weaker radiation force at the inlet of the trap to further facilitate microbubble entrance. A pulse sequence was generated that can switch between a long duration trapping waveform and short duration imaging waveform. High frame rate plane wave imaging was chosen for monitoring trapped microbubbles at 1 kHz. The working principle of the ultrasonic trap was explained and demonstrated in an ultrasound phantom by injecting SonoVue microbubbles flowing at 80 mL/min flow rate in a 3.5 mm diameter vessel.

\end{abstract}

\IEEEpeerreviewmaketitle

\section{Introduction}

Considerable research effort is being devoted to the development of microbubbles (MBs) with a therapeutic payload. A number of strategies have been developed to improve delivery of these agents, including targeting ligands, image guided acoustic release~\cite{Escoffre2013a}, sonoporation~\cite{McLaughlan2013}, magnetic targeting~\cite{Owen2015} and ultrasonic trapping with acoustic radiation force (ARF)~\cite{Raiton2012,Freear2013}. The aim of this study is to simultaneously image and trap MBs by using ARF with the ability of manipulating trapped MBs to increase the MB concentration at the region of interest inside blood vessels~\cite{Nie2018}.

Numerous biomedical applications use ARF including manipulation of cells in suspension, assessing viscoelastic properties of biological tissues, ablation therapy monitoring, targeted drug and gene delivery, molecular imaging, acoustical tweezers, and ultrasound-mediated thrombolysis~\cite{Sarvazyan2010,Xie2009}. The effect of primary and secondary radiation force on MBs are well studied~\cite{Leighton1994,Dayton1997,Doinikov1999a}. Effects of Bjerknes forces on MBs has already been evaluated and an acoustic trap was generated with two single element ultrasonic transducers~\cite{Yamakoshi2001}. This study investigates a new approach using a single array transducer for ultrasonically trapping MBs to increase the population at a desired location in a vessel phantom. The intended benefit of the ultrasonic trap is the increased efficacy in targeted drug delivery due to increased MB concentration around the region of interest. High frame rate ($>$ 1 kHz) plane wave imaging was used for monitoring the trapped and flowing MBs, where the imaging and the trapping sequences were interlaced. Working principle of the ultrasonic trap was demonstrated in a vessel phantom using SonoVue microbubbles.

\section{Materials and Methods}

\begin{figure}[!t]
	\centering
	\includegraphics[width = 80mm]{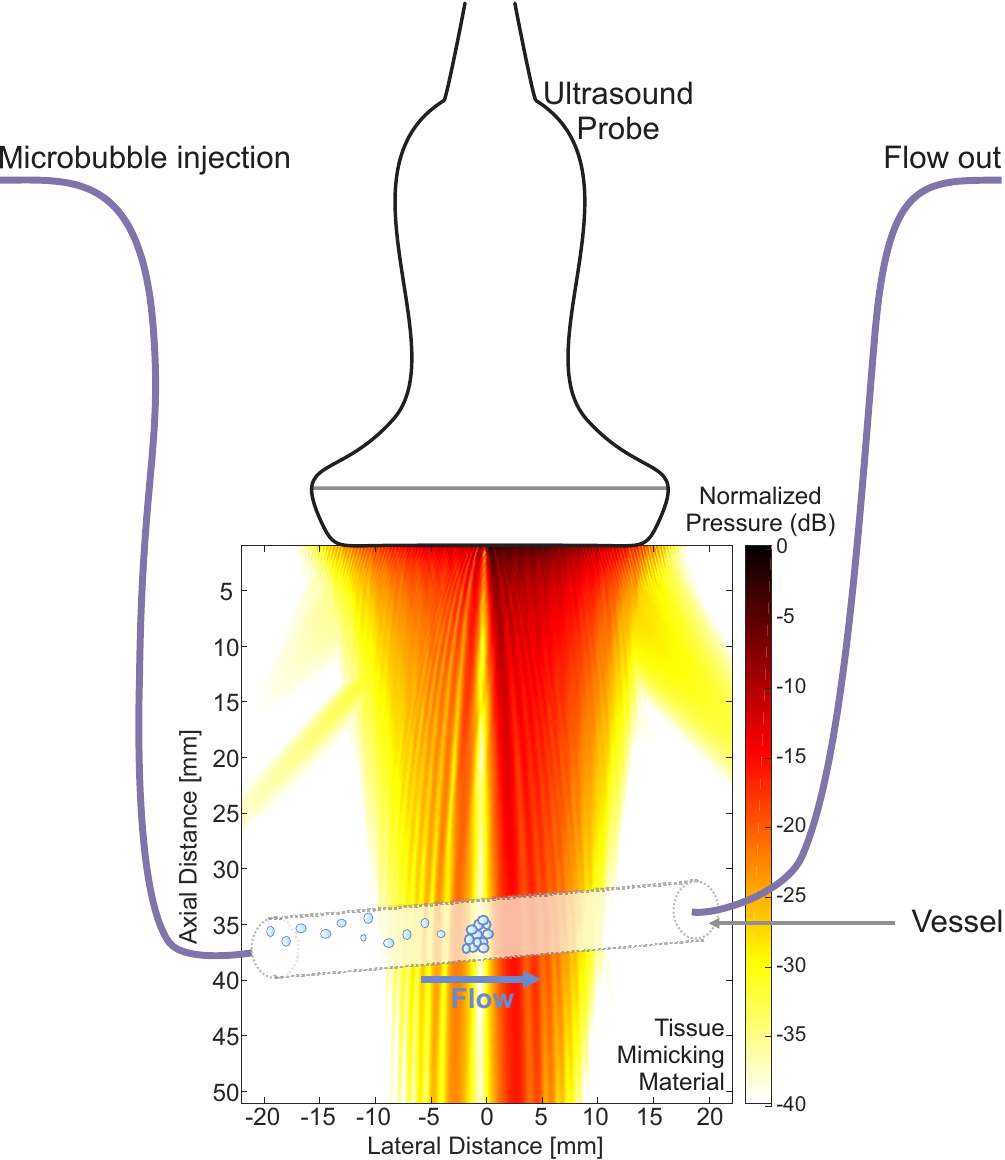}
	\caption{Illustration of the ultrasonic trap with an asymmetrical pressure field. Trapping region is located along 0~mm lateral axis. }
	\label{fig:Ultrasonic_Trap}
\end{figure}

The working principle of the ultrasonic trap is illustrated in Fig.~\ref{fig:Ultrasonic_Trap}. Trapping occurs at the low pressure region located along 0~mm lateral axis tightly wedged between two high pressure regions. The asymmetric trapping beam generates a weaker acoustic radiation force at the inlet (around -5 to -2 mm on lateral axis) and a stronger force at the outlet (around 1 to 5 mm on lateral axis) of the ultrasonic trap. This asymmetric shape facilitates MB entrance into the trap and allow for the pulsatile nature of flow.

\begin{figure*}[!t]
	\centering
	\includegraphics[width = 160mm]{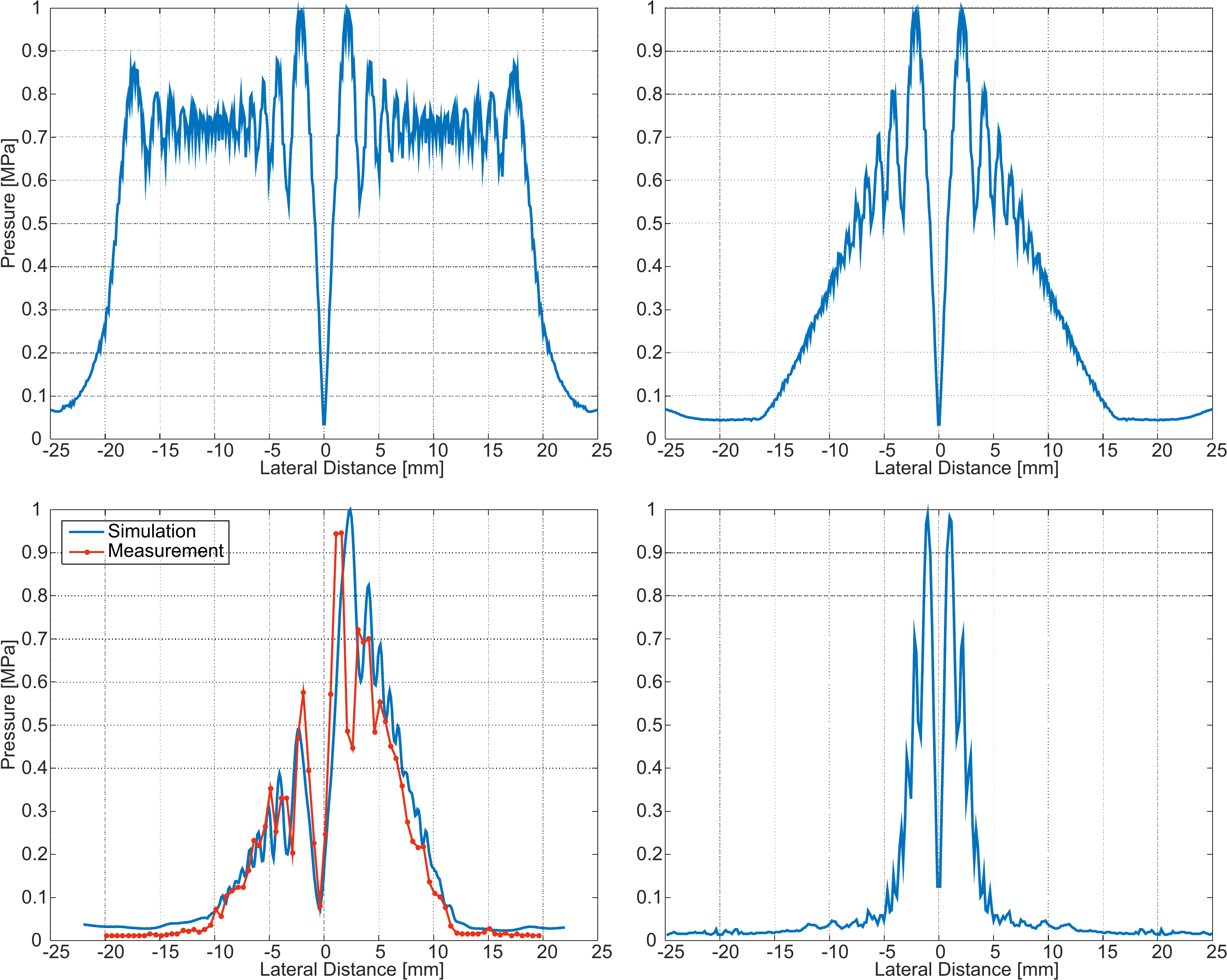}
	\caption{Figure shows the (blue) estimated and (red) measured pressure radiated from the transducer at 35~mm depth. A pressure null along 0~mm lateral axis is created for all excitation methods. (Top-left) Estimated pressure for two equal amplitude out of phase plane waves. (Top-right) Estimated pressure for two equal amplitude out of phase plane waves with apodization. (Bottom-left) Estimated and measured pressure for two out of phase asymmetric plane waves with apodization. This is the used method for ultrasonic trapping. (Bottom-right) Estimated pressure for two equal amplitude out of phase focused beams. This was the previously used method for ultrasonic trapping. }
	\label{fig:Ultrasonic_Trap_Pressure_meas_all}
\end{figure*}

\subsection{Acoustic Radiation Force}
The physical explanation of the ultrasonic trap is the acoustic radiation force acting on MBs. The ARF is generated due to the variation in the density of energy and momentum of the propagating waves~\cite{Sarvazyan2010}. A sudden pressure drop because of absorption, scattering or reflection can create such a force. For a travelling plane wave, the primary radiation force $F_1$ acting on a MB much smaller than the wavelength can be defined as
 \begin{equation}
	F_{1} = -   \left\langle  V(t) \nabla  P  \right\rangle 
\end{equation}
where $ \left\langle \right\rangle$ indicates time average, $V(t)$ is the MB volume, and $ \nabla P$ is the spatial pressure gradient. The primary radiation force for a MB with a resonant frequency $\omega_0$ can be expressed as~\cite{Dayton1997}
\begin{equation}
F_{1}=\frac{2\pi \,P^2 \,D \,R_0}{ \rho\,c\, \omega \,T} \frac{2\beta_{tot} / \omega}{((\omega_0 / \omega)^2 - 1)^2 + (2\beta_{tot} / \omega)^2}
\label{eq:prf}
\end{equation}
where $F_1$ drops with increasing driving frequency $\omega=2\pi f$.

Secondary radiation force originates from the pressure gradients in the re-radiated ultrasonic field from pulsating MBs~\cite{Harput2011}. The expansion and contraction of a MB generate a force that can attract or repel other MBs. This mutual interaction between the oscillating MBs can form stable clusters. The secondary radiation force can be expressed as~\cite{Dayton1997}
\begin{equation}
	\label{eq:linear_model}
	F_{2} = - \frac{2 \pi \,\rho }{9 \,d^{2}}   (P \,\omega)^{2} \,R_{1}^3 R_{2}^3 \, \epsilon_{1} \epsilon_{2} 
\end{equation}

Unlike primary radiation force, the secondary radiation force increases with increasing frequency. Therefore, an excitation frequency of 7 MHz, which is at the higher end of the transducer's bandwidth, was chosen to expedite MB aggregation. Another benefit of choosing a higher frequency is the smaller trap size and improved localization, where the trapping gap reduces with increasing frequency.

\subsection{Ultrasonic Trap}

The ultrasonic trap was generated by using the Leeds Ultrasound Array Research Platform 2 (UARP II) with capability of arbitrary excitation waveform control and ultra-fast image capture~\cite{Cowell2013,Smith2012,Smith2013a}. A 128 element linear medical imaging transducer (Prosonic, L3-8/40EP) was virtually divided into two sub apertures of 64 elements. This transducer was connected to the UARP II and an ultrasound pulse sequence was applied to each aperture with opposite phase polarity. The beams destructively interfere along the central axis of the transducer creating a pressure null that generates an acoustic trapping force on flowing MBs.

The design process of the ultrasonic trap is explained step by step in Fig.~\ref{fig:Ultrasonic_Trap_Pressure_meas_all} and compared with a similar acoustic trap generated by using focused beams. First, two sinusoidal tone bursts were transmitted with a central frequency of 7 MHz and duration of 500~$\mu$s from each aperture with opposite phase polarity. Fig.~\ref{fig:Ultrasonic_Trap_Pressure_meas_all}(Top-left) shows the estimated pressure radiated from the transducer at 35~mm depth. Two equal, out of phase, plane wave fields create a pressure null along 0~mm that can trap MBs. Assume MBs flowing in a blood vessel from negative to positive lateral direction as shown in Fig.~\ref{fig:Ultrasonic_Trap}. Drawback of this beam profile is the equally significant pressure gradients at centre of the trap and at -20~mm lateral axis, which prevents MBs from entering the trap. In order to reduce the pressure gradients located at -20~mm and +20~mm, a cosine shaped apodization window was applied over the array. Fig.~\ref{fig:Ultrasonic_Trap_Pressure_meas_all}(Top-right) shows the estimated pressure after applying apodization at the outer edges of the array. The resulting beam profile does not accommodate steep sided pressure gradients outside the trapping region. 

To further facilitate MB entrance to the trap, the intensity of the beam on the left aperture was reduced through the application of pulse width modulation~\cite{Smith2013a} and by keeping the apodization at the outer edges of the array, therefore generating a smooth beam shape with a weaker ARF at the inlet of the trap. Fig.~\ref{fig:Ultrasonic_Trap_Pressure_meas_all}(Bottom-left) shows the estimated beam profile at 35~mm depth. Pressure measurement performed at 35 mm depth in water with a differential membrane hydrophone matches with the predicted beam profile. Fig.~\ref{fig:Ultrasonic_Trap_Pressure_meas_all}(Bottom-right) shows the estimated pressure for a focused beam with opposite polarity. Again, the drawback of this beam profile is the equally significant pressure gradient around -3~mm, blocking MBs before entering the trap. Also both the inlet and the outlet of the trap have the same pressure gradient, which does not allow easy entrance into the trapping region.

\section{Results and Discussion}

Trapping and imaging sequences required two different beam profiles and signal durations that must be interleaved to achieve both. However with arterial blood velocity that can easily reach to 1 m/s, time spent gathering an imaging frame must be kept to a minimum. Linear imaging with a typical frame rate of 25 Hz requires 40 ms to acquire one frame, where MBs can travel 40 mm during the imaging sequence without the trapping beam. Some commercial systems can reach frame rates as high as 200 Hz while sector scan, however in 5 ms MBs can still travel 5 mm and flow out of the trapping region. By employing plane waves a frame rate of 10 kHz is achievable, where each frame acquisition requires 100~$\mu$s. During this time period without the trap, MBs can travel 100~$\mu$m, which is still within the trapping region. Therefore, the use of plane wave imaging is necessary while imaging and trapping MBs in arterial blood flow.

While performing the experiments, a syringe pump was driven with a flow rate of 80 mL/min to flow the diluted MB solution through the vessel inside the ultrasound phantom. Flow phantom was manufactured as described in \cite{Harput2013a}. The attenuation in tissue mimicking material was measured as $0.32$~dB/cm/MHz at 7~MHz, which reduced the peak negative pressure to 450~kPa with a MI of 0.17 inside the vessel at 35~mm depth. SonoVue microbubbles were diluted with deionized water by 1:500 to achieve similar concentrations to those observed in the human body. 

The flow rate of $Q=80$ mL/min corresponds to a mean fluid velocity of $V_{\rm mean}=140$ mm/s in flow phantom with a $d=3.5$~mm vessel, where maximum MB velocity can reach up to 280 mm/s. For such flow conditions in a Newtonian fluid, the wall shear rate $ \gamma = 8 V_{\rm mean} / d$ is expected to be 320~s$^{-1}$.

During the visualisation and trapping of SonoVue MB in the flow phantom, plane wave imaging was used with a frame rate of 1 kHz.  Although the UARP II system is capable of imaging at 10 kHz and above, it was not necessary due to slow flow rate inside the flow phantom. During the 1~ms imaging cycle and hence without the trap, MBs can travel up to 0.28~mm, while still remaining within the trapping region. In either trapping or imaging mode the whole 128 elements of the array was employed to ensure optimal trapping control and the highest quality images.

\begin{figure}[!t]
	\centering
	\includegraphics[width = 88mm]{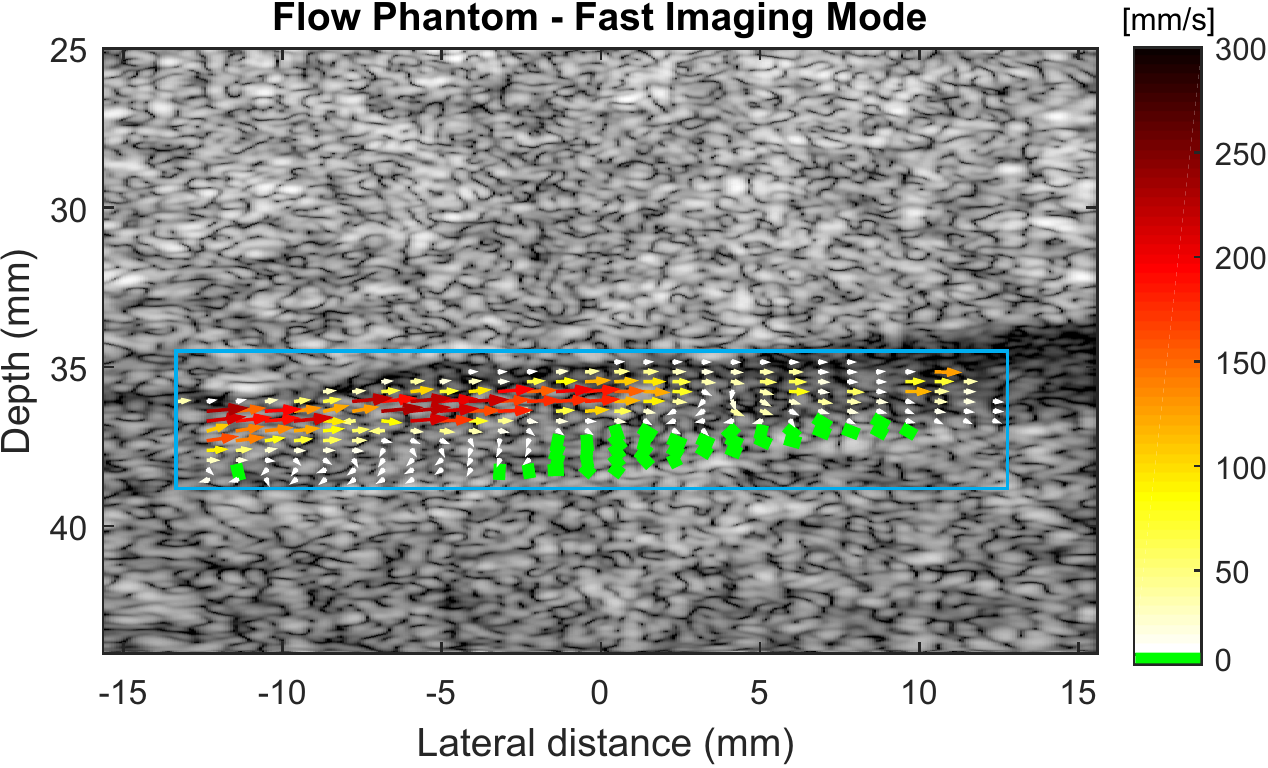}
	\caption{Velocity profile of MBs overlaid on B-mode image captured with 1 kHz frame rate. Trapped MBs are represented with green dots (0-5~mm/s) and flowing MBs are represented with coloured arrows according to their speed and trajectories. }
	\label{fig:Trap_velocity_profile}
\end{figure}

A speckle tracking algorithm was used to detect the trapped MBs (green dots) and flowing MBs (coloured arrows) within the flow phantom as shown in Fig.~\ref{fig:Trap_velocity_profile}~\cite{Nie2016}. The MBs travelling with a velocity lower than 5~mm/s are considered to be trapped for a flow rate of 80~mL/min. All MBs are affected differently by the ARF. For example, two MBs excited above or below their resonance frequencies will result in an attractive force, while if one is below and the other is above resonance, the net force will be repulsive. However, the resonance frequency of an aggregated MB decreases and they all eventually start attracting each other after the initial aggregation onset.

Some off-resonance MBs can still escape the ultrasonic trap, however the velocity of the MBs drops significantly after the trapping region due to the strong ARF. The density of the trapped MBs are the highest around back vessel wall due to the residual primary radiation. When a big MB cluster is formed, the drag force acting against the MBs increases. Although the drag force eventually beats the trapping force for a large cluster of MB, because of the golf-ball shape, a trapped MB cluster size can reach up to 10-20~$\mu$m that corresponds to hundreds of MBs. When the trapping sequence was stopped, the trapped MBs disperse and normal contrast agent flow returns.  

\section{Conclusions}

This study showed the viability of using acoustic radiation force to form an ultrasonic MB trap in a tissue mimicking phantom. The trapping force was created by generating spatial pressure variations with predesigned ultrasound beam patterns using a linear array imaging transducer. A custom designed asymmetric and apodized beam profile was developed which can retain MBs in the trapping zone and is resilient to pulsatile clinical flow rates. MB monitoring was performed with the same transducer utilizing high frame rate plane wave imaging, interlaced with the trapping sequence. Through the generation of a trapping force directly opposing the flow, the ultrasonic trap was able to halt MBs subject to wall shear rates of up to 320~s$^{-1}$ at a mechanical index of 0.17.


For drug loaded MBs, ultrasonic trapping can potentially increase the drug volume at a particular location while continuously monitoring with high speed plane wave imaging. Although, it is hard to differentiate between tissue and trapped stationary MBs with flow imaging methods, various signal processing techniques such as bispectral analysis can be employed to achieve this separation~\cite{Harput2012}. Further work will focus on electronically shifting the trapping region to manipulate MBs, quantifying the cluster size and identify the location of trapped MBs with ultrasound imaging, and destruction of the MB clusters.

\section*{Acknowledgment}     
The authors gratefully acknowledge funding from the Engineering and Physical Sciences Research Council (UK) via grant no. EP/K030159/1, EP/I000623/1 and the Leverhulme fellowship (ECF-2013-247).

\newpage

\bibliography{BuBBle,Ultrasound}       
\bibliographystyle{IEEEtran}

\end{document}